\documentclass{aa}  

\usepackage{graphicx}
\usepackage{txfonts}
\usepackage{color}

\begin{document}

   \title{The application of the Mid-IR radio correlation to the
     $\hat{G}$ sample and the search for advanced extraterrestrial
     civilisations}

\titlerunning{The application of the Mid-IR radio correlation to the
     $\hat{G}$ sample}


   \author{M. A. Garrett\inst{1,2} }

   \institute{ASTRON, Netherlands Institute for Radio Astronomy,
     Postbox 2, 7990AA Dwingeloo, The Netherlands \and
Leiden Observatory, Leiden University, P.O. Box 9513, 2300 RA Leiden, The Netherlands.\\
              \email{garrett@astron.nl}
             }

   \date{Received June XX, 2015; accepted August 11, 2015}

   \abstract{Wright et al. 2014a have embarked on a search for
     advanced Karadashev Type III civilisations via the compilation of
     a sample of sources with extreme mid-IR emission and colours. The
     aim is to furnish a list of candidate galaxies that might harbour
     an advanced Kardashev Type III civilisation - in this scenario,
     the mid-IR emission is then primarily associated with waste heat
     energy by-products. I apply the Mid-IR radio correlation to this
     $\hat{G}$ (Glimpsing Heat from Alien Technology) sample, a
     catalogue of 93 candidate galaxies compiled by
     Griffith et al. 2015. I demonstrate that the mid-IR and radio
     luminosities are correlated for the sample, determining a
     k-corrected value of $q_{22}=1.35\pm0.42 $.  
     By comparison, a similar measurement for 124
     galaxies drawn from the First Look Survey (FLS) has
     $q_{22}=0.87\pm0.27$. The statistically significant difference of
     the mean value of $q_{22}$ for these two samples, taken together with their
     more comparable FIR properties, suggests that the $\hat{G}$
     sample shows excessive emission in the Mid-IR. 
     The fact that the G-HAT sample largely
     follows the Mid-IR radio correlation, strongly suggests the vast 
     majority of these sources are associated with galaxies in
     which natural astrophysical processes are dominant. 
     This simple application of the mid-IR radio
     correlation can substantially reduce the number of false
     positives in the $\hat{G}$ catalogue, since galaxies occupied by
     advanced Kardashev Type III civilisations would be expected to
     exhibit very high values of $q$. Indeed I identify 9 outliers in the
     sample with $q_{22} > 2$ of which at least 3 have
     properties that are relatively well explained via standard
     astrophysical interpretations e.g. dust emission associated
     with nascent star formation and/or nuclear activity from a heavily
     obscured AGN.  The other outliers have not been
     studied in any great detail, and are deserving of further
     observation. I also note that the comparison of resolved Mid-IR and
     radio images of galaxies on sub-galactic (kpc) scales can also be
     useful in identifying and recognising artificial mid-IR emission
     from less advanced intermediate Type II/III civilisations.
     Nevertheless, from the bulk properties of the $\hat{G}$ sample, I
       conclude that Kardashev Type-III civilisations are either very
       rare or do not exist in the local Universe. 
}


   \maketitle
%

\section{Introduction}

In a series of papers, \citet{Wright2014a} and \citet{Wright2014b} 
have presented a detailed description of our current ability to detect the
signature of advanced Kardashev Type III civilisations \citep{Kardashev1964}
via the prominent waste heat signature they are expected to produce.
Type III civilisations are defined by \citet{Kardashev1964} as those
capable of harnessing the stellar energy supply of a galaxy 
($\sim 10^{38}$ Watts). Indeed previous studies (e.g. \citet{Carrigan2009}), suggest that
constructs such as Dyson Spheres \citep{Dyson1960} will radiate
most of their waste heat energy at Mid-Infrared (MIR) wavelengths, corresponding to
temperatures of $\sim 100-600K$. 

Wright et al. have embarked on a novel project $\hat{G}$
(Glimpsing Heat from Alien Technology) based on the results {\bf of} an all-sky
Mid-Infrared (MIR) survey conducted by the WISE mission \citep{Wright2010}. 
In particular, \citet{Griffith2015} have recently produced a list of 93
sources (from an original sample of 100000 resolved WISE detections)
that exhibit both extreme mid-IR emission and mid-IR colours. If the 
radiation measured by WISE is interpreted as waste heat emission from
an advanced civilisation, the source sample includes
galaxies reprocessing more than 25\% of their starlight into the
Mid-IR ({\it i.e.} $\gamma > 0.25$ in the formalism of \citet{Wright2014b}). While some of
these sources are well known ({\it e.g.} Arp 220), the majority have not
been individually studied in any great
detail. One significant problem with the $\hat{G}$
approach is the large number of false positives expected in the sample
- in particular, there are many
ways in which emission in the mid-IR can arise via natural
astrophysical processes e.g. the reprocessing of starlight or AGN
radiation by dust. 

One way of identifying {\it bona fide} Type-III
civilisations is to identify outliers in well determined scaling laws for galaxies
e.g. the Tully-Fisher relation \citet{Annis1999}. I 
argue here, that the Infra-red radio correlation can also be used in a similar
way. The original Infrared radio correlation is a fundamental
relation for galaxies (\citet{vanderkruit1971}, \citet{Helou1985},
  \citet{Condon1992}, \citet{Yun2001}),  covering 
at least 5 orders of magnitude in luminosity, holding
over a wide range of different redshifts, and extending well into the
FIR/Mid-Infrared and sub-millimetre domains (\citet{Carilli1999}, \citet{Garrett2002}, \citet{Elbaz2002},
\citet{Ivison2002}, \citet{Appleton2004}). Studies of the correlation
usually quote $q$, the logarithm of the ratio of the IR to radio flux
densities (luminsoities), the latter typically being measured at 1.4
GHz ($\lambda 20$cm). 
K-corrected values of $q$ vary from $\sim 2.3$ in the FIR (60-100
micron) to $\sim 1$ in the
Mid-IR (24 micron), and reflect the evolvution of the bolometric spectral energy
distribution (SED) of a galaxy across the radio, NIR, MIR and FIR domains. The  
correlation is tightest for star forming galaxies
but also applies to many other galaxy types, including radio quiet AGN
(\citet{Roy1998}). 

The physical explanation for the tightness of the correlation is that
both the non-thermal radio emission and the thermal IR emission are
related to mechanisms driven by massive star formation. For galaxies
in which the bulk of the Mid-IR emission is associated with waste heat
processes, there is no obvious reason why artifical radio emission
would be similarly enhanced. While the continuum radio emission level
might increase through the use of advanced communication systems,
the amount of waste energy deposited in the radio domain is likely to
be many orders of magnitude less than that expected at Mid-IR
wavelengths. As a consequence, I propose to apply the Mid-IR radio correlation
to the 93 $\hat{G}$ sources presented in Table 9 of
\citet{Griffith2015}). In particular, galaxies that are associated
with Type-III civilisations, should appear as outliers in the mid-IR
radio correlation with extremely high values of $q$. It should be noted that
this deviation is opposite in sense to other frequent outliers {\it i.e.}
radio-loud AGN with systematically low values of $q$).


In this paper, I calculate values of $q$ for the  $\hat{G}$ sample,
identifying interesting outliers that are deserving of further study. In
section 2 I introduce details of the sample and the auxiliarly radio
and IR data gathered together. Section 3 displays the main
results, and these are discussed further in section 4. Section 5 presents the
main conclusions of the paper with a suggestion for further work. 


\section{The $\hat{G}$ sample and auxiliary data}

\citet{Griffith2015} have identified $\sim100000$ resolved sources
detected by WISE, and located above the galactic plane
($b > 10^{\circ}$).  By avoiding the Milky Way and similarly dense
regions of sky such as the LMC, and by applying various colour
criteria, \citet{Griffith2015} eliminate confounding objects such as
galactic stars, diffuse nebular emission and other stellar
artefacts. The resulting cleaned catlogue of $\sim 31000$ extended red
objects is therefore biased to include nearby galaxies that are
prominent MIR sources with extreme colours. In addition,
\citet{Griffith2015} apply the AGENT methodology
\citep{Wright2014b}) to identify 93 sources
that have MIR colours consistent with $\gamma > 0.25$ where $\gamma$
is the fraction of starlight re-emitted in the MIR as waste heat
products, as modelled by \citet{Wright2014b}'s AGENT analysis of the
four WISE observing bands (3.4, 4.6, 12 and 22 microns).

I have cross-matched the $\hat{G}$ sample of 93 sources (see
\citet{Griffith2015}, Table 9) with the NRAO/VLA Sky Survey (NVSS) 1.4
GHz (20cm) radio catalogue \citep{Condon1998}. I have restricted our study
to the 92 sources that fall within the NVSS survey area and have measured
redshifts. Since the sources
are all resolved by WISE, they are mostly local systems with the
median redshift being 0.028. The highest redshift source in
the sample has $z = 0.14525$. I compare the
$\lambda 20$cm radio emission with the WISE (band 4) 22 micron data since
for local galaxies this band is is dominated by continuum
emission. The remaining WISE bands are more sensitive to (PAH) spectral
features. Redshifts for the sources, together with IRAS 60 and 100
micron flux densities (where available), were extracted from a search
of the NASA/IPAC Extragalactic Database (NED) and SIMBAD. The WISE
magnitude system was converted to Janskys following
\citet{Wright2010}. The main data are presented in Table 1. 

\section{Determining k-corrected values of $q$ for the $\hat{G}$ sub-sample}

Following \citet{Appleton2004} I define
$q_{22} = log(S_{22\mu}/S_{20cm})$ where $S_{22\mu}$ and $S_{20cm}$
are the source flux densities measured by WISE and NVSS at wavelengths
of $22\mu$m (WISE band 4) and $20$cm. Since the majority of the source
sample are typically located in the local universe, the values of
$q_{22}$ derived are relatively insensitive to any (reasonable)
k-correction. Nevertheless, a k-correction has been applied to the
analysis presented here in order to ensure consistency with other
author's results obtained for higher-z samples. For the radio, I adopt
a k-correction factor of $(1+z)^{+0.7}$ following
\citet{Appleton2004}. For the mid-IR corrections, and in particular
$q_{22}$, I assume an M82 like SED as presented in
\citet{Sturm2000}. Over the limited redshift range associated with the
$\hat{G}$ sample ($22-26\mu$m), the k-correction is also well
modeled by a power-law: $(1+z)^{-2.45}$.

Fig.1 presents a plot of the k-corrected $22\mu$ mid-IR luminosity ($L_{22\mu}$,
W/Hz) against the 20cm radio luminosity ($L_{20cm}$, W/Hz) for the
$\hat{G}$ sample (red dots). The data clearly show a strong correlation between the
Mid-IR and radio luminsoities. A formal fit to the observed
correlation yields: \\

$log(L_{22\mu m})=(0.76\pm0.09)log(L_{20cm}) +(6.65\pm0.37). $
\\

For a subset of the $\hat{G}$ sample (37 of the 92 sources) identified by \citet{Griffith2015} as
galaxies so excluding AGN, Sy 1 \&2 etc)  I find (similarly): \\

$log(L_{22\mu m})=(0.79\pm0.15)log(L_{20cm}) +(6.08\pm0.33). $
\\

\begin{figure}
   \centering
   \includegraphics[width=9.5cm]{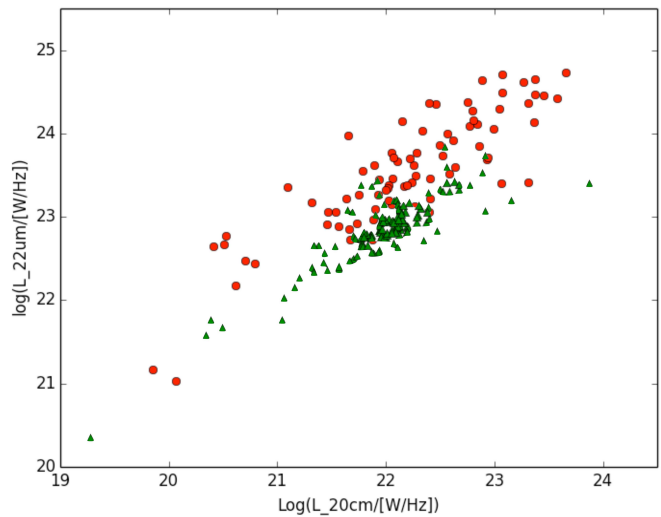}
      \caption{The k-corrected $22\mu$ mid-IR luminosity ($L_{22\mu}$,
W/Hz), plotted against the 20cm radio luminosity ($L_{20cm}$, W/Hz) for the
$\hat{G}$ (red circles) and FLS (green triangles) sample. 
              }
         \label{Fig1}
   \end{figure}

In addition, a k-corrected value of $q_{22}=1.35\pm0.42$ is derived for the full
sample. By comparison a value of $q_{22}=1.40\pm0.34$ is also derived, again
for the same subset of 37 sources identified as
galaxies. Table 1, presents the values of $q$ derived for the full source sample,
including upper limits for the 10 sources that are below the 2.5mJy
detection threshold of NVSS. The table is presented in order of
decreasing $q_{22}$. 

As far as I can ascertain, $q_{22}$ has not yet been derived for any
other source samples observed by WISE thus far. \citet{Appleton2004} derived a zero-redshift value for
  $q_{24}=0.84\pm0.28$ for a source sample derived from Spitzer
  observations of the First Look Survey (FLS). In Fig.1 I {\bf also} present
  the data for the $\hat{G}$ sample together with a 
  k-corrected plot of $L_{22\mu}$ vs $L_{20cm}$
  (W/Hz) for 124 galaxies located in the FLS that I also identify in the
  WISE all-sky catalogue with sources of known redshift, such that $z <
  0.2$ \citep{Marleau2007}. As expected, the FLS sample (plotted as triangles) clearly show a strong correlation between
  the mid-IR and radio luminosities with a spread that is smaller than
  the $\hat{G}$ galaxy sample. The slope of the best linear fit to the FLS
  data is similar to that seen in the $\hat{G}$ sub-sample.  A formal fit 
to the observed correlation in the FLS galaxy sample yields: \\

$log(L_{22\mu m})=(0.70\pm0.07)log(L_{20cm}) +(7.36\pm0.22)$
\\

I derive a k-corrected
  $q_{22}=0.87\pm0.27$ for this FLS sub-sample. Considering the small
  wavelength difference, this is plainly
  consistent with the values of $q_{24}$ derived by
  \citet{Appleton2004}.

\begin{figure}
   \centering
\includegraphics[scale=0.329]{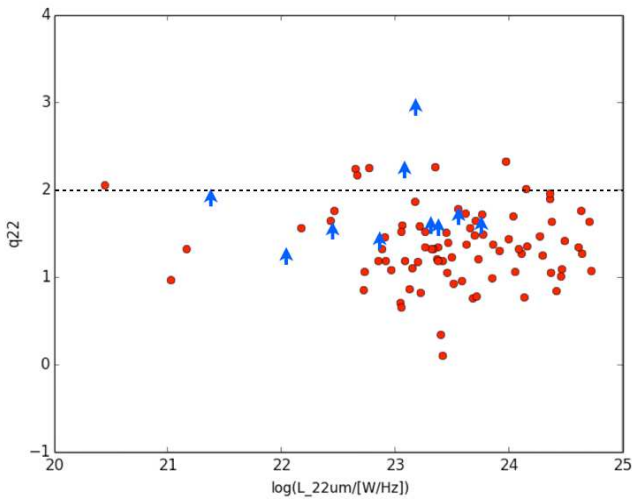}
      \caption{{\bf The k-corrected values of  $q_{22}$ plotted against the
        $22\mu$ mid-IR luminosity ($L_{22\mu}$,
W/Hz) for the
$\hat{G}$ sample (red filled circles). Above the dashed line, lie the
9 outliers from the sample with $q_{22} > 2 $. Sources not detected in
the NVSS radio survey show lower limits for $q_{22}$, and are presented as arrows.}
              }
         \label{Fig2}
   \end{figure}

  Inspection by eye of Fig.1, suggests that the
      mean difference between the ($\hat{G}$) and FLS samples
      ($q_{22}=1.35\pm0.42$ vs $q_{22}=0.87\pm0.27$) may be
      significant. Indeed, applying Student's T-test \citep{Student1908} to the data, show the difference between the means
      (0.48) to be statistically significant at the 95\% confidence level (with the
      two tailed P value < 0.0001, t(204)=9.99, a pooled variance of
      0.45 and a 95\% confidence
      interval of the mean difference ranging between lower and upper
      limits of 0.3852 to 0.5748).   

   $q_{FIR}$ was {\bf also} determined for
   a subset of the $\hat{G}$ sample (those sources with both 60 and 100
   micron flux densities) finding $q_{FIR}=2.45+/-0.39$. This is
   consistent with values found in much larger local galaxy samples e.g.
   $q_{FIR}=2.34\pm0.1$. (Yun, Reddy \& Condon 2001.). 

  Systematic underestimates of the radio flux density do not seem to
  be a major factor in our analysis, despite the extended nature of
  the sources, and the limited uv-coverage afforded by the NVSS
  snapshot observations. In particular, the fitted sizes of the
  sources are typically less than the NVSS synthesised beam (i.e. <
  45 arcsec). I conclude that the higher value of $q$ for the
  $\hat{G}$ sub-sample is a physical characteristic of the source
  sample, and this has its origins in enhanced mid-IR emission rather
  than some systematic radio defecit. 

Fig. 2 presents a plot of $q_{22}$ against the $22\mu$ mid-IR luminosity ($L_{22\mu}$,
W/Hz) of the $\hat{G}$ sample. Outlying sources with $q_{22} > 2$ are
positioned above the dashed line. Details of the same sources are also presented
towards the top of Table 1. 

\section{Discussion} 

Both the Mid-IR and radio luminsoties of sources in the $\hat{G}$ sub-sample
are strongly correlated with each other. This clearly demonstrates
that the source sample as a whole follows the well-established Mid-IR radio correlation
associated with natural astrophysical processes such as massive star
formation. 
One interesting, though perhaps not unexpected feature of the $\hat{G}$
sample is that the mean value of 
$q_{22}$ appears to be statistically different and indeed larger than
that determined for the FLS
sample. The fact that the values of $q_{FIR}$ are
comparable between the samples, suggests that this probably reflects a
systematic excess in Mid-IR emission assocociated with the  $\hat{G}$
sample, rather than a defecit of the sample in the radio domain.

Large  values of $q_{22}$ would be expected for
systems dominated by Kardashev Type III civilisations, and this makes
the outliers in the $\hat{G}$ sample of particular
interest. Sources with  $q_{22}>2$ 
 (i.e. those lying $>1.5\sigma$ off of the $\hat{G}$
  sample mean or $>4\sigma$ off of the FLS sample mean), 
include: MCG+02-60-017, IC 342, ESO 400-28, NGC 814, NGC 4747, NGC 5253, UGC
3097, NGC 4355 and NGC 1377. Astrophysical explanations for high
values of $q$ in the mid-IR include (i) very young star forming systems in
which the synchrotron radio component is not fully established or (ii) 
obscured AGN that heat nuclear dust to relatively warm temperatures.
Of the sources with $q_{22}>2$ presented here, NGC 1377, NGC 4355 and
IC 342 have been studied in some depth. \citet{Altao2012} favour an interpretation for NGC 1377
in which the prominent molecular outflow is driven by a young AGN
embedded in a dust enshrouded nucleus. By comparison, IC 342 is a
nearby face-on barred spiral galaxy with a central nuclear starburst,
fed by bar-driven gas inflow \citep{Schinnerer2003}.  The compact
nucleus of NGC 4355 (also known as NGC 4418) also harbours an
extremely rich and dusty molecular environment but it is unclear
whether a compact starburst or an AGN (or some combination of both)
power the strong mid-IR compnent \citep{Varenius2014}. These three
examples, are probably typical representations of the range of type
that dominate the $\hat{G}$ sample with large values of $q_{22}$.

\section{Conclusions and next steps}

In this paper, I have demonstrated that the IR-radio correlation can
be employed as a useful diagnostic in distinguishing
between Mid-IR emission produced by natural astrophyiscal processes and
that generated by artificial means {\it e.g.} the waste heat energy associated with Type III
civilisations. In particular, galaxies dominated by Type III
civilisations should present themselves as extreme outliers to the
Mid-IR radio correlation with values of $q_{22}$ > 2. In this way, the
mid-IR radio correlation can be used to eliminate false
positives from the $\hat{G}$ sample, and identify those systems that deserve further detailed study. 

The observations presented here, demonstrate that the $\hat{G}$ sample
of 93 sources (\citet{Griffith2015}, Table 9) typically follow the
IR-radio correlation. I suggest that the vast majority, if not all of
these sources present mid-IR
emission associated with natural astrophysical processes. Those sources with
$q_{22}$>2 that are have not yet been widely studied in the literature
do deserve further investigation however. Nevertheless, I 
conclude that the vast majority, if not all of the sources in the
$\hat{G}$ sample do not obviously
harbour Karadashev Type III civilisations, and that therefore such
  civilisations are either extremely rare in the local universe or do
  not exist. 

Finally, it should be noted that the IR-radio correlation is also known to hold on
sub-galactic scales (e.g. \citet{Murphy2006}. A comparison of
resolved Mid-IR and radio images of nearby galaxies on kpc scales can also be useful
in identifying artificial mid-IR emission from advanced civilisations
that lie between the Type II and Type III types. While
\citet{Wright2014a} venture that Type III civilisations should emerge rapidly from
Type IIs, it might be that some specific galactic localities are
preferred - see for example \citet{Cirkovic2006} or are to be best avoided e.g. the galactic centre. 
A comparison of the resolved radio and mid-IR structures can therefore
also be
relevant to future searches of waste heat associated with advanced
civilisations.  

\begin{acknowledgements}
      Part of this work was supported by an IBM Faculty Award. This
      research has made use of the NASA/IPAC Extragalactic Database
      (NED) which is operated by the Jet Propulsion Laboratory,
      California Institute of Technology, under contract with National
      Aeronautics and Space Administration. This research has made use
      of the SIMBAD database, operated at CDS, Strasbourg, France. 
This research has made use of the NASA/ IPAC Infrared Science Archive,
which is operated by the Jet Propulsion Laboratory, California
Institute of Technology, under contract with the National Aeronautics
and Space Administration. I would like to thank the referee for 
comments and suggestions, that have helped to strengthen the paper. 

\end{acknowledgements}

%
%

\clearpage 
\begin{table*}
\caption{\label{t7}The $\hat{G}$ sample studied in this paper,
  including values (or upper limints) of $q_{22}$.}
\centering
\tiny
\begin{tabular}{llccccccl}
\hline\hline
RA (J2000) & Dec (J2000) & $S_{20cm}$ & WISE-4 (mag) & $q$ & z & $\gamma$ & SIMBAD Type & Name  \\
\hline
03 36 39.05 & -20 54 06.8 & < 2.5 & 1.5 & > 2.88 & 0.005921 & 0.5 &
                                                                    GiG & NGC 1377 \\
12 26 54.61 & -00 52 39.1 & 40.8 & -0.15 & 2.326 & 0.007048 & 0.85 & Sy2 & NGC 4355 \\
04 35 48.45 & +02 15 29.6 & 3.9 & 2.54 & 2.263 & 0.012014 & 0.52 & G & UGC 3097 \\
13 39 55.96 & -31 38 24.4 & 84.7 & -0.73 & 2.248 & 0.001345 & 0.60 & AGN & NGC 5253 \\
12 51 45.54 & +25 46 28.5 & 7.4 & 1.92 & 2.244 & 0.003966 & 0.53 & IG & NGC 4747 \\
02 10 37.63 & -15 46 24.2 & 5.0 & 2.54 & 2.164 & 0.005405 & 0.54 & G & NGC 814 \\
20 28 25.49 & -33 04 20.5 & < 2.5 & 3.29 & > 2.156 & 0.01228 & 0.32 & G & ESO 400-28 \\
03 46 48.35 & +68 05 46.5 & 190.7 & -1.12 & 2.054 & 7.7e-05 & 0.49 & GiG & IC 342 \\
23 47 09.20 & +15 35 48.3 & 9.2 & 2.2 & 2.008 & 0.026105 & 0.65 & Sy2 & MCG+02-60-017  \\
00 36 52.44 & -33 33 16.8 & 26.8 & 1.17 & 1.963 & 0.020558 & 0.66 & AGN & ESO 350-38 \\
22 52 34.71 & +24 43 49.4 & 7.3 & 2.67 & 1.9 & 0.04193 & 0.60 & AGN & Mrk 309 \\
18 16 03.19 & +47 37 05.4 & 3.2 & 3.74 & 1.862 & 0.01706 & 0.39 & G & 2MASX J18160312+4737056 \\
10 25 08.18 & +17 09 14.1 & < 2.5 & 4.14 & > 1.829 & 0.002512 & 0.58 & IG & NGC 3239 \\
10 01 25.94 & +15 46 12.2 & 44.0 & 1.14 &  1.776 & 0.007895 & 0.34 & G & NGC 3094 \\
05 55 42.61 & +03 23 31.8 & 32.5 & 1.52 & 1.763 & 0.002669 & 0.72 & H2G & UGCA 116 \\
04 34 00.03 & -08 34 44.9 & 137.1 & -0.07 & 1.756 & 0.015851 & 0.63 & AGN & NGC 1614 \\
10 38 33.62 & -07 10 14.4 & 66.6 & 0.8 & 1.733 & 0.007283 & 0.61 & G & IC 630 \\
05 43 23.63 & +54 00 44.2 & 6.9 & 3.24 & 1.715 & 0.027019 & 0.51 & G & 2MASX J05432362+5400439 \\
13 02 20.39 & -15 45 59.0 & 34.8 & 1.55 & 1.702 & 0.016745 & 0.53 & GiG & MCG-02-33-099 X \\
05 16 46.24 & -12 20 59.4 & 6.9 & 3.47 & 1.651 & 0.00636 & 0.26 & G & 6dFGS gJ051646.2-122100 \\
08 35 38.40 & -01 14 07.1 & 2.7 & 4.38 & 1.645 & 0.0438 & 0.36 & G & 2MASX J08353838-0114072 \\
17 38 01.51 & +56 13 25.9 & 5.8 & 3.51 & 1.634 & 0.06519 & 0.45 & Sy2 & 2MASX J17380143+5613257 \\
20 57 24.32 & +17 07 38.5 & 43.2 & 1.43 & 1.632 & 0.035 & 0.70 & G & IRAS F20550+1655SE \\
17 34 31.80 & +47 13 01.7 & < 2.5 & 4.54 & > 1.621 & 0.03901 & 0.29 & G & 2MASX J17343177+4713010 \\
15 48 13.36 & -24 53 09.6 & 6.9 & 3.59 & 1.593 & 0.0139 & 0.39 & EmG & ESO 515-7 \\
19 46 05.40 & +64 08 50.1 & 5.5 & 3.84 & 1.585 & 0.018786 & 0.26 & G & 2MASX J19460544+6408494 \\
05 56 52.46 & -05 23 03.8 & 3.1 & 4.55 & 1.564 & 0.007749 & 0.26 & LSB & 2MASXI J0556522-052308 \\
23 36 14.11 & +02 09 17.9 & 65.8 & 1.24 & 1.559 & 0.00935 & 0.49 & AGN & NGC 7714 \\
10 54 16.74 & -39 40 19.3 & 5.6 & 3.98 & 1.524 & 0.016611 & 0.31 & G & ESO 318-23 \\
14 28 37.03 & -39 48 44.1 & < 2.5 & 4.81 & > 1.521 & 0.03298 & 0.28 & IG & ESO 326-24 \\
18 25 52.75 & +37 52 41.6 & < 2.5 & 4.74 & > 1.52 & 0.05492 & 0.32 & IR & IRAS 18241+3750 \\
12 09 13.87 & +26 52 37.4 & 2.1 & 5.0 & 1.518 & 0.03447 & 0.26 & G & LEDA 38612 \\
06 43 39.31 & -27 12 17.6 & 7.0 & 3.74 & 1.514 & 0.02353 & 0.27 & GiG & 2MASX J06433935-2712180 \\
 02 45 06.40 & -02 07 27.7 & < 2.5 & 4.86 & > 1.496 & 0.03656 & 0.29 & G & 2MFGC 2186 \\
13 45 47.40 & +70 04 45.9 & 8.3 & 3.58 & 1.493 & 0.032012 & 0.37 & G & 2MASX J13454733+7004455 \\
19 57 51.88 & -32 21 28.2 & 13.0 & 3.15 & 1.48 & 0.0239 & 0.35 & EmG & 6dFGS gJ195751.9-322128 \\
00 54 04.02 & +73 05 05.7 & 113.8 & 0.84 & 1.473 & 0.015823 & 0.54 & G & MCG+12-02-001  \\
05 01 47.35 & -18 10 00.8 & < 2.5 & 5.02 & > 1.463 & 0.01319 & 0.30 & G & NGC 1739 \\
06 45 40.95 & +43 34 07.5 & 3.3 & 4.72 & 1.453 & 0.01982 & 0.27 & G & 2MASX J06454097+4334069 \\
13 20 21.98 & -23 32 25.9 & 8.1 & 3.71 & 1.435 & 0.0448 & 0.35 & EmG & 2MASX J13202200-2332256 \\
22 44 58.08 & -01 46 00.4 & 13.3 & 3.16 & 1.417 & 0.06262 & 0.41 & G & 2MASX J22445816-0145589 \\
13 15 03.51 & +24 37 07.8 & 30.8 & 2.45 & 1.4 & 0.013049 & 0.61 & Q? & IC 860 \\
02 41 43.24 & +45 46 26.6 & 7.5 & 3.99 & 1.372 & 0.032736 & 0.38 & G & 2MASX J02414325+4546272 \\
19 41 12.82 & +63 05 42.9 & 5.0 & 4.37 & 1.37 & 0.05267 & 0.31 & G & 2MASX J19411289+6305430 \\
 18 52 22.44 & -29 36 20.7 & 16.0 & 3.18 & 1.354 & 0.042355 & 0.37 & EmG & 6dFGS gJ185222.4-293621 \\
19 12 27.31 & -29 02 35.7 & 7.3 & 4.1 & 1.349 & 0.025629 & 0.35 & EmG & ESO 459-7 \\
15 18 06.13 & +42 44 44.8 & 50.3 & 1.96 & 1.347 & 0.0405 & 0.60 & LIN & IRAS F15163+4255NW \\
00 42 50.05 & -36 52 43.2 & 6.7 & 4.22 & 1.34 & 0.02374 & 0.30 & G & 6dFGS gJ004250.1-365241  \\
06 32 26.08 & -24 32 08.3 & < 2.5 & 5.29 & > 1.34 & 0.02416 & 0.30 & G & ESO 490-11 \\
15 23 09.66 & -39 34 48.2 & 7.2 & 4.17 & 1.327 & 0.02541 & 0.34 & G & 2MASX J15230967-3934481 \\
10 20 50.93 & -17 18 59.4 & 5.2 & 4.52 & 1.323 & 0.0293 & 0.32 & EmG & MCG-03-27-005  \\
03 04 45.12 & +07 47 39.2 & 2.3 & 5.42 & 1.321 & 0.02682 & 0.27 & G & 2MASX J03044511+0747394 \\
12 15 39.36 & +36 19 35.1 & 33.7 & 2.59 & 1.321 & 0.000977 & 0.59 & SBG & NGC 4228 \\
02 08 05.42 & -29 14 32.7 & 6.4 & 4.19 & 1.321 & 0.0636 & 0.41 & G & 6dFGS gJ020805.4-291433 \\
02 55 59.96 & +47 48 19.3 & 19.0 & 3.16 & 1.302 & 0.031332 & 0.37 & G & MCG+08-06-022  \\
00 18 50.88 & -10 22 36.6 & 42.4 & 2.37 & 1.275 & 0.027 & 0.52 & EmG & MCG-02-01-051  \\
15 34 57.25 & +23 30 11.5 & 326.3 & 0.19 & 1.272 & 0.018116 & 0.79 & SyG & Arp 220 \\
15 02 53.22 & +16 55 08.4 & 18.7 & 3.4 & 1.226 & 0.021223 & 0.33 & rG & NVSS J150253+165507 \\
01 20 02.63 & +14 21 42.5 & 49.8 & 2.24 & 1.251 & 0.031555 & 0.49 & LIN & MCG+02-04-025  \\
00 08 20.57 & +40 37 55.9 & 6.9 & 4.44 & 1.211 & 0.046148 & 0.54 & Sy2 & 2MASX J00082041+4037560 \\
09 23 38.22 & -25 16 34.9 & 14.4 & 3.74 & 1.203 & 0.0214 & 0.32 & GiP & 6dFGS gJ092338.2-251635 \\

\hline
\end{tabular}
\tablefoot{}
\end{table*} 

\clearpage 

\begin{table*}
\caption{\label{t7}Table 1 continued: The $\hat{G}$ sample studied in this paper,
  including values (or upper limints) of $q_{22}$.}
\centering
\tiny
\begin{tabular}{llccccccl}
\hline\hline
RA (J2000) & Dec (J2000) & $S_{20cm}$ & WISE-4 (mag) & $q$ & z & $\gamma$ & SIMBAD Type & Name  \\
\hline

04 28 51.45 & +69 34 47.1 & 9.3 & 4.26 & 1.192 & 0.016168 & 0.33 & G & 2MASX J04285125+6934469 \\
04 47 38.41 & -17 26 01.8 & 11.7 & 4.02 & 1.192 & 0.013299 & 0.39 & G & ESO 552-5 \\
14 07 36.99 & +16 01 21.6 & 4.7 & 4.98 & 1.185 & 0.02768 & 0.25 & G & 2MASX J14073693+1601212 \\
09 35 48.86 & -29 19 55.6 & 4.1 & 5.08 & 1.185 & 0.0431 & 0.32 & EmG & ESO 434-13 \\
12 53 24.16 & -23 45 45.6 & 3.1 & 5.37 & 1.184 & 0.0475 & 0.31 & EmG & 6dFGS gJ125324.2-234546 \\
18 10 41.37 & +25 07 23.2 & 9.7 & 4.23 & 1.178 & 0.02213 & 0.38 & G & 2MASX J18104135+2507238 \\
05 47 38.65 & -10 35 52.8 & < 2.5 & 5.76 & > 1.169 & 0.01154 & 0.25 & G & 6dFGS gJ054738.7-103552 \\
05 43 23.63 & +54 00 44.2 & 6.9 & 4.77 & 1.103 & 0.027019 & 0.56 & G & 2MASX J05432362+5400439 \\
10 59 18.14 & +24 32 34.6 & 57.0 & 2.44 & 1.098 & 0.042876 & 0.50 & LIN & 2XMM J105918.1+243234 \\
09 39 30.15 & +06 26 13.0 & 5.7 & 5.04 & 1.082 & 0.02459 & 0.25 & rG & NVSS J093929+062610?2MAS \\
13 44 42.10 & +55 53 13.3 & 144.7 & 1.51 & 1.072 & 0.03734 & 0.55 & Sy2 & Mrk 273 \\
04 55 10.80 & +05 35 12.6 & 7.5 & 4.81 & 1.065 & 0.01671 & 0.38 & G & 2MASX J04551070+0535126 \\
23 26 10.59 & -30 31 06.1 & 10.5 & 4.29 & 1.065 & 0.0641 & 0.28 & EmG & 6dFGS gJ232610.6-303106 \\
16 36 00.57 & +10 13 40.4 & 4.1 & 5.09 & 1.055 & 0.14525 & 0.43 & G & 2MASX J16360060+1013396 \\
15 00 29.00 & -26 26 49.2 & 37.8 & 3.08 & 1.053 & 0.01751 & 0.65 & H2G & 2MASX J15002897-2626487 \\
12 13 46.00 & +02 48 40.3 & 23.3 & 3.54 & 1.008 & 0.0731 & 0.55 & LIN & LEDA 39024 \\
22 54 51.03 & +37 42 20.8 & 3.8 & 5.49 & 0.993 & 0.091137 & 0.28 & G & 2MASX J22545096+3742205 \\
02 59 41.29 & +25 14 15.0 & 33.1 & 3.49 & 0.969 & 0.001265 & 0.54 & G & NGC 1156 \\
01 16 07.20 & +33 05 21.7 & 75.4 & 2.57 & 0.959 & 0.016044 & 0.54 & Sy2 & NGC 449 \\
05 44 21.56 & -13 53 11.9 & 10.5 & 4.7 & 0.932 & 0.0403 & 0.31 & G & 2MASX J05442151-1353116 \\
17 34 29.01 & -04 05 41.7 & 6.2 & 5.44 & 0.87 & 0.0363 & 0.35 & G & 6dFGS gJ173428.9-040542 \\
05 15 21.42 & -26 28 17.1 & 20.6 & 4.25 & 0.855 & 0.012809 & 0.30 & EmG & ESO 486-39 \\
12 49 30.16 & -11 24 03.4 & 73.4 & 2.79 & 0.842 & 0.047826 & 0.59 & Sy2 & IRAS 12468-1107 \\
16 18 33.99 & +13 24 25.9 & 4.6 & 5.85 & 0.818 & 0.04953 & 0.28 & G & 2MASX J16183392+1324253 \\
22 21 49.97 & +39 50 24.0 & 7.8 & 5.31 & 0.779 & 0.069725 & 0.29 & G & 2MASX J22215143+3950240 \\
00 34 43.48 & -00 02 26.6 & 57.0 & 3.25 & 0.774 & 0.042529 & 0.54 & Sy2 & 2MFGC 403 \\
02 56 09.76 & -15 39 43.5 & 10.9 & 5.04 & 0.756 & 0.05877 & 0.33 & G & 6dFGS gJ025609.8-153946 \\
11 24 02.72 & -28 23 15.4 & 52.1 & 3.6 & 0.711 & 0.01374 & 0.43 & Sy2 & IRAS 11215-2806 \\
11 45 43.59 & -11 47 12.6 & 33.3 & 4.21 & 0.655 & 0.018463 & 0.29 & EmG & 6dFGS gJ114543.6-114712 \\
10 50 52.15 & +01 09 44.2 & 32.7 & 4.94 & 0.343 & 0.03952 & 0.29 & GiG & IC 649S \\
13 08 42.02 & -24 22 57.8 & 474.5 & 2.72 & 0.103 & 0.014 & 0.50 & Sy2 & PKS 1306-241 \\

\hline
\end{tabular}
\tablefoot{}
\end{table*} 

\end{document}